\documentclass[conference]{IEEEtran}

\ifCLASSINFOpdf
\else
\fi

\usepackage{graphicx}
\graphicspath{{./pdf/}{./jpeg/}{./images/}}
   \DeclareGraphicsExtensions{.pdf,.jpeg,.jpg,.png}
   
\usepackage{eurosym}
\usepackage{color,soul}
\usepackage[numbers]{natbib}

\begin{document}
\title{Battling the Digital Forensic Backlog\\through Data Deduplication}

\author{\IEEEauthorblockN{Mark Scanlon}
\IEEEauthorblockA{School of Computer Science,\\
University College Dublin,\\
Ireland.\\
E-mail: mark.scanlon@ucd.ie}
}

\maketitle

\begin{abstract}
In recent years, technology has become truly pervasive in everyday life. Technological advancement can be found in many facets of life, including personal computers, mobile devices, wearables, cloud services, video gaming, web-powered messaging, social media, Internet-connected devices, etc. This technological influence has resulted in these technologies being employed by criminals to conduct a range of crimes -- both online and offline. Both the number of cases requiring digital forensic analysis and the sheer volume of information to be processed in each case has increased rapidly in recent years. As a result, the requirement for digital forensic investigation has ballooned, and law enforcement agencies throughout the world are scrambling to address this demand. While more and more members of law enforcement are being trained to perform the required investigations, the supply is not keeping up with the demand. Current digital forensic techniques are arduously time-consuming and require a significant amount of man power to execute. This paper discusses a novel solution to combat the digital forensic backlog. This solution leverages a deduplication-based paradigm to eliminate the reacquisition, redundant storage, and reanalysis of previously processed data.
\end{abstract}

\IEEEpeerreviewmaketitle

\section{Introduction}
The number of cases requiring digital forensic analysis has significantly increased in recent years. Moreover, each individual case may require the acquisition and analysis of digital evidence from a multitude of media to gain a comprehensive understanding of the case in question. These sources can include computer equipment, mobile devices, file synchronisation servers, cloud servers, web service data, social media, wearables, navigation equipment, instant messaging, email, etc. The sheer volume of potentially evidence-rich data alongside the complexity involved in accessing and acquiring the data from such a variety of sources leads to extended processing time per case. As a result, digital evidence backlogs are commonplace in law enforcement agencies throughout the world \cite{casey2009investigation, mislan2010growing, lillis2016challenges}. These backlogs in local, state and national police forces commonly reach one to two years, and in some extreme cases can exceed four years \cite{lillis2016challenges}. During this time, potentially pertinent evidence is sitting in an evidence locker unimaged and unanalysed. As a result, case detectives may be left waiting for prolonged periods of time for digital forensic processing to be performed in order to unearth a lead that will progress their direction of investigation \cite{hitchcock2016tiered}.

The acquisition speed of digital evidence from its source can be limited by a number of influencing factors. A number of these factors can be alleviated through infrastructure investment by the digital forensics laboratory, e.g., workstation speed, evidence storage write speeds, local network transfer speed, etc. However, a significant number of influencing factors are outside the forensic laboratories' control, e.g., read speed of the target device's storage, remote server speeds, Internet bandwidth, etc. Little can be done by a forensic investigator to expedite the acquisition speed of data from remote, third-party sources, e.g., website-hosting servers, file synchronisation servers, etc. Acquiring data from these sources relies on the third-party's cooperation through available APIs or specialist cloud-evidence recovery tools \cite{scanlon2014leveraging}.

During the traditional evidence acquisition phase, any hardware device being investigated will be connected via a write-blocker to a forensic workstation. Subsequently, an exact bit-by-bit copy of the entire storage medium (hard disk, flash storage, memory card, etc.) is typically acquired using industry-standard software, such as EnCase or FTK Imager \cite{quick2014impacts}. This copy can be stored on the local machine for instant processing or stored on a network-attached storage device in the forensic laboratory. This process can take in the order of minutes for small storage devices (such as small USB keys, camera memory cards, etc.) and up to tens of hours for large multi-terabyte disks or network-attached storage devices. The limiting factor for the acquisition speed is normally the read speed of the device being imaged. During this prolonged period of time, the forensic workstation is typically entirely taken up with this single procedure -- and this is before the evidence analysis phase begins.

This paper discusses a database-driven, deduplicated data approach to alleviating the digital forensic backlog. Data deduplication involves the usage of a centralised storage system to store a single copy of each object added. When an object needs to be stored in the system, a check is first performed to see if that object already exists in the storage system; determining its addition or not. This check is performed comparing the object's cryptographic hash value with the hash values of all files currently in the storage system. A similar technique has previously been employed by cloud-based file synchronisation tools to save on storage and bandwidth costs, such as Dropbox. 








\section{Digital Forensic Backlog}

The digital forensic backlog is a common problem encountered by law enforcement agencies throughout the world. Despite the best efforts of digital forensic laboratories, backlogs in the order of six months to one year are commonplace \cite{casey2009investigation, grier2015rapid}, frequently reaching two years \cite{lundy2015need}, and can reach up to four years in the extreme \cite{lillis2016challenges}. Due to the requirement for expert analysis exceeding the current capability in these agencies, the queue of casework grows. As an indication of the increasing volume of information being processed, in 2013 the FBI Regional Computer Forensics Laboratory reported to have processed 5,973TB of data from 7,273 examinations (up 40\% when compared with 2011) \cite{al2016suspect}. \citet{quick2014impacts} outline three main factors contributing to the digital forensic backlog:

\begin{enumerate}
\item An increase in the number of cases whereby digital forensic analysis is required by the investigation.
\item An increase in the number of devices that are seized for digital forensic processing per case.
\item The volume of potentially pertinent data stored on each device seized is also increasing.
\end{enumerate}

The sheer volume of cases, coupled with the volume of data to be processed per case, is set to increase further in the foreseeable future \cite{lillis2016challenges, garfinkel2010digital}. This ``volume challenge'' has been long identified as one of the greatest threats to digital forensics \cite{grier2015rapid}. Without a scalable, extensible solution to this volume challenge, the backlog can be reasonably assumed to worsen into the future.

\section{Requirements for Proposed System}

While a number of the requirements for the proposed system are consistent with any digital forensic tool (forensically sound evidence handling, built-in audit trail, reproducibility of process, etc.), this section outlines a number of additional requirements for the system providing numerous benefits over the traditional approach.

\subsection{Breaking the Acquisition Performance Wall}

Using current evidence acquisition techniques, entire disk images are taken from the original data and subsequently the analysis is performed on these images. The bottleneck during this acquisition phase is typically the read speed of the source device. Data deduplication in isolation cannot expedite this process, i.e., the act of calculating each artefact's hash value requires that the entirety of that artefact be read into memory. However, if the investigation is particularly time-sensitive, an option should exist for the investigator to introduce an acceptable element of risk to the deduplication process for the benefit of speed. This risk involves the hashing of a small chunk of the artefact (as opposed to the entire artefact) to be used for the deduplication process. As a result, an entire hard disk image could be acquired and reconstructed server-side faster than the sequential copying of the entire disk.

In a remote evidence acquisition scenario, e.g., uploading the device image to a Digital Forensics as a Service (DFaaS) service, the bottleneck will likely be the upload speed of the investigator's Internet connection. The greater the number of artefacts eliminated using the deduplication method, the faster the overall acquisition time will be, e.g., an entire disk image could be reconstructed server-side faster than it takes to send the complete image over the Internet.

\subsection{Expedited Digital Forensic Processing}
A number of techniques have been employed to streamline the current digital forensic process, including efficient workflow management \cite{braekt2016workflow}, DFaaS \cite{vanBaar2014S54, van2015digital}, triage \cite{shaw2013practical, hitchcock2016tiered} and automation \cite{al2016suspect, james2015automated}. However, significant resources are being wasted with the current processing model; both in terms of the computer and manpower overheads. 

From a computational perspective, significant resources are wasted in the reacquisition, duplicated storage, and reanalysis of digital evidence. This wastage both occurs across cases and across digital forensic laboratories. Significant investment is made in the infrastructure necessary for each digital forensic laboratory to perform its task. From a budgetary perspective, centralising the processing of digital forensic evidence can provide significant cost savings for law enforcement agencies. This ensures that each unit has always-on access to the latest technologies and techniques at the centralised resource. Departments with little funding for digital forensic infrastructure will no longer be at a disadvantage in investigating their cases.

With limited staffing, maximising the productivity of digital investigators is crucial to help alleviate the backlog. Significant time is wasted in waiting for the acquisition of large hard drives. In smaller forensic laboratories, the prolonged time needed for the acquisition phase can greatly hamper the throughput of the laboratory. This is due to forensic workstations being singularly utilised by the acquisition task.

Data deduplication is expected to eliminate a significant proportion of the necessary data to be processed. \citet{watkins2009teleporter} found that data deduplication resulted in savings of 30-95\% in their testing of a similar deduplication-based system. The authors expect that a full implementation of their system should see 50-70\% deduplication from typical real-world hard drives.

\begin{figure*}
\centering
\includegraphics[width=0.75\textwidth]{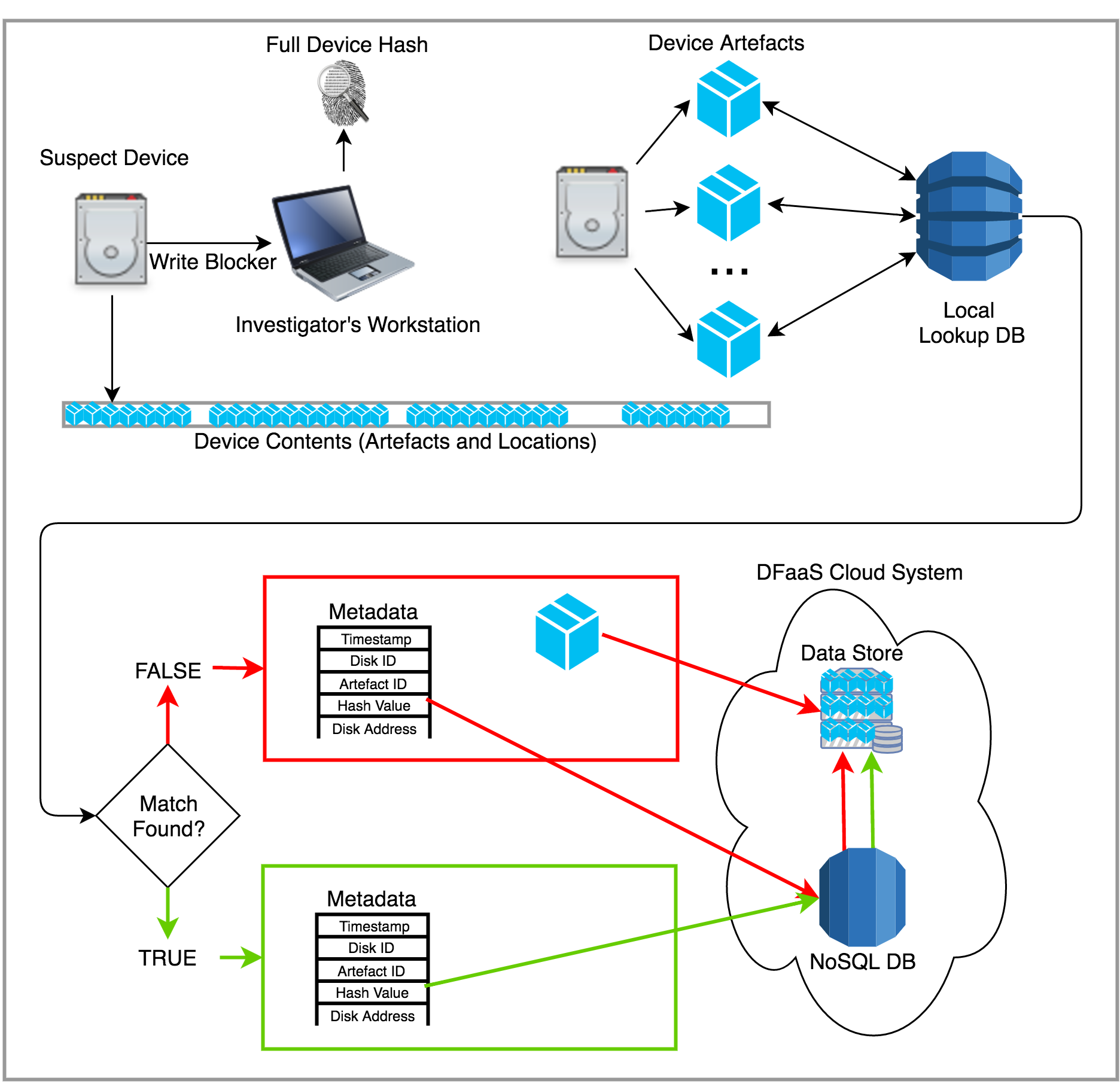}
\caption{Disk Image Acquisition Phase}
\label{fig:acquisition}
\end{figure*}

\subsection{Collaborative Examination and Sharing}
Due to the decentralised nature of the Internet, police departments investigating online crime quickly find themselves relying on international collaboration to progress their investigations. Existing international policing agreements, e.g., Europol and Interpol, enable cross-border collaboration in digital forensic cases. Current approaches for this collaboration can often involve local police departments shipping the physical devices to requesting international agencies. Moving to a cloud-based solution for digital forensic processing and sharing should facilitate easier cooperation on a local, national and international level. Having a centralised shared resource among collaborative countries can greatly expedite this entire process. 

\section{Proposed Solution}

It is envisioned that the proposed solution outlined in this section would be implemented in a centralised datacentre or preferably on a cloud-based infrastructure. The premise of the system involves moving away from the individual digital forensic laboratories to a centralised processing model -- ideally built on a DFaaS model, such as that described by \citet{vanBaar2014S54}. This centralised, shared model facilities a number of advantages outlined in Section~\ref{advantages}. 

The traditional approach to data acquisition involves the connecting of the suspect device to a forensic workstation using a write-blocker (ensuring no inadvertent modification). Subsequently a bit-by-bit copy of the original data is taken \cite{casey2009investigation}. This copy is then verified as a true copy of the original through the comparison of the data's hash values, as can be seen in the top left of Figure~\ref{fig:acquisition}. This image is then typically transferred to a network-attached storage device for later analysis.

Using the proposed system, the target storage medium would be acquired piece-by-piece. In common with the traditional approach, the input storage device would be connected to the forensic workstation using a suitable write-blocker. It would then be analysed artefact-by-artefact (files, file fragments, uninitialised/unallocated space, etc.). Each of these artefact's hash values would be compared against a local lookup database, as shown in Figure~\ref{fig:acquisition}. In the scenario entire device acquisition, this local database would represent all the files contained in the main centralised database. In an evidence whitelisting/blacklisting scenario, this database need only represent all of the pertinent artefacts to that investigation type. Regardless of whether the artefact was acquired previously or not, all associated metadata from this acquisition are stored in the database.
 
In terms of system reliability, the weak link of this data deduplicated model becomes the deduplication itself. In the event of unrecoverable data loss, e.g., hard disk failure, fire, etc., deduplication can prove unreliable. Of course, in a real-world implementation of the proposed system, a suitable off-site backup procedure for the file store and the database would be put in place to ensure their recoverability in the event of disaster.

\begin{figure*}[t]
\centering
\includegraphics[width=\textwidth]{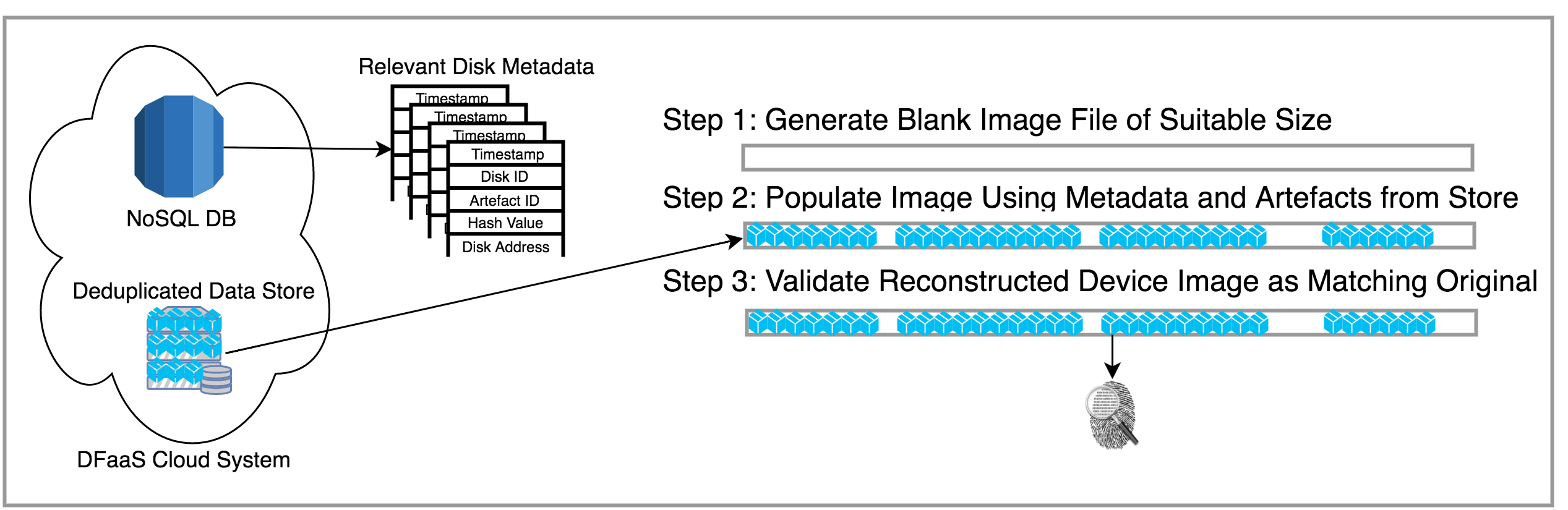}
\caption{Complete Disk Image Reconstruction Phase}
\label{fig:recon}
\end{figure*}

\subsection{Elimination of Redundant Analysis}
In addition to the metadata recorded during the acquisition phase, each artefact analysed by digital forensic experts can be easily categorised in the centralised database. The categorisation of benign operating system files, application executables, and commonly encountered files can result in the elimination of computational and expert processing time in their reanalysis. Known pertinent/incriminating artefacts can also be quickly identified and automatically flagged to the investigator at the earliest stage possible during the investigation. 

Another benefit of this database-driven approach is the ability to quickly and easily create a variety of ``incriminating'' datasets, which can be used for efficient device/suspect blacklisting/whitelisting. These sets of known incriminating hash values can be updated on-the-fly whenever needed ensuring the most up-to-date information possible.

\subsection{Disk Image Reconstruction}

To begin entire disk reconstruction, first a blank target disk image would be created. Then the metadata associated with that specific acquisition would be retrieved from the database. Subsequently, each artefact associated with that particular acquisition would be gathered and inserted into its corresponding location, as shown in Figure~\ref{fig:recon}. In this manner, an entire disk image can be reconstructed using artefacts that may have originally been acquired, indexed and stored from previous cases. 

Eliminated data identified during the acquisition phase of a new device can be instantly added to the target image server-side. This on-the-fly disk image reconstruction can result in a complete disk image being more quickly available when compared to the traditional approach.

\subsection{Advantages over Traditional Approach}
\label{advantages}
Focusing on the goal of battling the digital forensic backlog, switching to the proposed database-driven deduplication system results in a number of advantages including:

\begin{itemize}
\item \textit{Reduced Storage Requirements} -- The proposed deduplicated system need only store a single copy of each unique artefact (file, file fragment, slackspace/unallocated space) discovered from all investigations conducted using the system. This results in significant cost savings or more being achievable with existing infrastructure.
\item \textit{Less Data to be Acquired per Device} -- The computational overhead in the reacquisition and storage of previously discovered content is eliminated. Operating system files, application files, previously encountered benign and incriminating data, etc., need not be reacquired from the device.
\item \textit{Reduced Bandwidth} -- In the scenario of remote evidence acquisition, such as the acquisition of digital evidence over the Internet \cite{scanlon2010online}, the reduction in the volume of data being transferred will result in vastly expedited remote acquisitions.
\item \textit{Automated Blacklisting/Whitelisting} -- Employing the proposed deduplicated system alongside sufficient categorisation of artefacts, the easy creation and maintenance of a database of all incriminating files is facilitated. Such a database could have a high update frequency, ensuring new files are added to the list as they are discovered and analysed. This database could then be used for automated blacklisting/whitelisting of suspect machines. Blacklisting/whitelisting at the earliest point in an investigation can prioritise the subsequent processing towards relevant suspects/devices and can eliminate the processing of non-pertinent devices.
\item \textit{Simultaneous Acquisition and Processing} -- Due to the fact that duplicated artefacts may have already been processed by another expert using the system, preliminary automated processing of known data can take place.
\item \textit{Reduction in Manual Analysis} -- As cases are processed by digital forensic experts using the system, each artefact encountered could be optionally marked as benign, i.e., irrelevant to any future investigation, or incriminating, i.e., any future encountering of that artefact can be automatically flagged.
\item \textit{Focused Expert Analysis} -- In the scenario where a sufficient volume of cases and associated evidence has been analysed and processed by digital forensic experts, many of the artefacts typically encountered on common devices will be eliminated from reprocessing. The investigator's time is saved from the reanalysis of known content and quickly focused on the new content discovered on the device in question. This can also aid at the digital forensic triage stage (\cite{hitchcock2016tiered, rogers2006computer}) with the investigator only being presented with unknown content or incriminating content at the earliest point possible in the investigation.
\item \textit{Point-in-time Reconstruction} -- High frequency acquisition of a single suspect machine is also possible using the proposed system. Due to the fact that only the filesystem changes would need to be acquired, the footprint of each acquisition would be minimal. Point-in-time reconstruction would then be possible from the deduplicated store.
\item \textit{The Bigger, the Better} -- As a juxtaposition with the ``volume challenge'', the more investigations performed using the proposed system increases the impact of each of the aforementioned advantages listed above. The more artefacts in the system, more duplicates would be discovered in future investigations. Once a significant file store has been populated, a number of future research areas emerge surrounding automated evidence analysis through feature extraction from the existing dataset.
\end{itemize}

\section{Concluding Remarks}
The digital evidence backlog is already a hindrance frequently encountered in modern policing. Contributing factors to the backlog include the volume of cases, the volume of data, limited resourcing, limited manpower, alongside an overly arduous digital forensic process. These factors are set to continue to negatively influence the throughput of digital forensic laboratories into the foreseeable future. While there are a number of complementary solutions to combat the backlog, e.g., increased digital forensic laboratories budgets, increased expert training, increased resources, etc., a paradigm shift is needed to streamline and future-proof the digital forensic process. This paper discusses the advantages of a centralised data deduplication system over the current digital forensic process. The proposed system is capable of alleviating some of the backlog through the elimination of duplicated efforts (both computational and personnel), while providing a number of enhancements to the functionality available with the traditional alternative.

\bibliographystyle{IEEEtranN}
\bibliography{intech,../bib/paper}

\begin{thebibliography}{18}
\providecommand{\natexlab}[1]{#1}
\providecommand{\url}[1]{#1}
\csname url@samestyle\endcsname
\providecommand{\newblock}{\relax}
\providecommand{\bibinfo}[2]{#2}
\providecommand{\BIBentrySTDinterwordspacing}{\spaceskip=0pt\relax}
\providecommand{\BIBentryALTinterwordstretchfactor}{4}
\providecommand{\BIBentryALTinterwordspacing}{\spaceskip=\fontdimen2\font plus
\BIBentryALTinterwordstretchfactor\fontdimen3\font minus
  \fontdimen4\font\relax}
\providecommand{\BIBforeignlanguage}[2]{{%
\expandafter\ifx\csname l@#1\endcsname\relax
\typeout{** WARNING: IEEEtranN.bst: No hyphenation pattern has been}%
\typeout{** loaded for the language `#1'. Using the pattern for}%
\typeout{** the default language instead.}%
\else
\language=\csname l@#1\endcsname
\fi
#2}}
\providecommand{\BIBdecl}{\relax}
\BIBdecl

\bibitem[Casey et~al.(2009)Casey, Ferraro, and Nguyen]{casey2009investigation}
E.~Casey, M.~Ferraro, and L.~Nguyen, ``Investigation delayed is justice denied:
  proposals for expediting forensic examinations of digital evidence,''
  \emph{Journal of forensic sciences}, vol.~54, no.~6, pp. 1353--1364, 2009.

\bibitem[Mislan et~al.(2010)Mislan, Casey, and Kessler]{mislan2010growing}
R.~P. Mislan, E.~Casey, and G.~C. Kessler, ``The growing need for on-scene
  triage of mobile devices,'' \emph{Digital Investigation}, vol.~6, no.~3, pp.
  112--124, 2010.

\bibitem[Lillis et~al.(2016)Lillis, Becker, O'Sullivan, and
  Scanlon]{lillis2016challenges}
D.~Lillis, B.~Becker, T.~O'Sullivan, and M.~Scanlon, ``{Current Challenges and
  Future Research Areas for Digital Forensic Investigation},'' 05 2016.

\bibitem[Hitchcock et~al.(2016)Hitchcock, Le-Khac, and
  Scanlon]{hitchcock2016tiered}
B.~Hitchcock, N.-A. Le-Khac, and M.~Scanlon, ``Tiered forensic methodology
  model for digital field triage by non-digital evidence specialists,''
  \emph{Digital Investigation}, vol.~16, no.~S1, pp. 75--85, 03 2016,
  proceedings of the Third Annual DFRWS Europe.

\bibitem[Scanlon et~al.(2014)Scanlon, Farina, Le~Khac, and
  Kechadi]{scanlon2014leveraging}
M.~Scanlon, J.~Farina, N.-A. Le~Khac, and M.-T. Kechadi, ``{Leveraging
  Decentralisation to Extend the Digital Evidence Acquisition Window: Case
  Study on BitTorrent Sync},'' pp. 85--99, September 2014.

\bibitem[Quick and Choo(2014)]{quick2014impacts}
D.~Quick and K.-K.~R. Choo, ``Impacts of increasing volume of digital forensic
  data: A survey and future research challenges,'' \emph{Digital
  Investigation}, vol.~11, no.~4, pp. 273--294, 2014.

\bibitem[Grier and Richard(2015)]{grier2015rapid}
J.~Grier and G.~G. Richard, ``Rapid forensic imaging of large disks with
  sifting collectors,'' \emph{Digital Investigation}, vol.~14, pp. S34--S44,
  2015.

\bibitem[Lundy(2015)]{lundy2015need}
G.~Lundy, ``The need for distributed lab resources for state local tribal and
  territorial law enforcement agencies,'' Ph.D. dissertation, UTICA College,
  2015.

\bibitem[Al~Fahdi et~al.(2016)Al~Fahdi, Clarke, Li, and Furnell]{al2016suspect}
M.~Al~Fahdi, N.~Clarke, F.~Li, and S.~Furnell, ``A suspect-oriented intelligent
  and automated computer forensic analysis,'' \emph{Digital Investigation},
  vol.~18, pp. 65--76, 2016.

\bibitem[Garfinkel(2010)]{garfinkel2010digital}
S.~L. Garfinkel, ``Digital forensics research: The next 10 years,''
  \emph{digital investigation}, vol.~7, pp. S64--S73, 2010.

\bibitem[In~de Braekt et~al.(2016)In~de Braekt, Le-Khac, Farina, Scanlon, and
  Kechadi]{braekt2016workflow}
R.~In~de Braekt, N.-A. Le-Khac, J.~Farina, M.~Scanlon, and M.-T. Kechadi,
  ``{Increasing Digital Investigator Availability through Efficient Workflow
  Management and Automation},'' pp. 68--73, 04 2016.

\bibitem[van Baar et~al.(2014)van Baar, van Beek, and van Eijk]{vanBaar2014S54}
R.~van Baar, H.~van Beek, and E.~van Eijk, ``Digital forensics as a service: A
  game changer,'' \emph{Digital Investigation}, vol. 11, Supplement 1, pp. S54
  -- S62, 2014, proceedings of the First Annual DFRWS Europe.

\bibitem[van Beek et~al.(2015)van Beek, van Eijk, van Baar, Ugen, Bodde, and
  Siemelink]{van2015digital}
H.~van Beek, E.~van Eijk, R.~van Baar, M.~Ugen, J.~Bodde, and A.~Siemelink,
  ``Digital forensics as a service: Game on,'' \emph{Digital Investigation},
  vol.~15, pp. 20--38, 2015.

\bibitem[Shaw and Browne(2013)]{shaw2013practical}
A.~Shaw and A.~Browne, ``A practical and robust approach to coping with large
  volumes of data submitted for digital forensic examination,'' \emph{Digital
  Investigation}, vol.~10, no.~2, pp. 116--128, 2013.

\bibitem[James and Gladyshev(2015)]{james2015automated}
J.~I. James and P.~Gladyshev, ``Automated inference of past action instances in
  digital investigations,'' \emph{International Journal of Information
  Security}, vol.~14, no.~3, pp. 249--261, 2015.

\bibitem[Watkins et~al.(2009)Watkins, McWhorte, Long, and
  Hill]{watkins2009teleporter}
K.~Watkins, M.~McWhorte, J.~Long, and B.~Hill, ``Teleporter: An analytically
  and forensically sound duplicate transfer system,'' \emph{digital
  investigation}, vol.~6, pp. S43--S47, 2009.

\bibitem[Scanlon and Kechadi(2009)]{scanlon2010online}
M.~Scanlon and M.-T. Kechadi, ``Online acquisition of digital forensic
  evidence,'' in \emph{Proceedings of International Conference on Digital
  Forensics and Cyber Crime (ICDF2C 2009)}.\hskip 1em plus 0.5em minus
  0.4em\relax Albany, New York, USA: Springer, September 2009, pp. 122--131.

\bibitem[Rogers et~al.(2006)Rogers, Goldman, Mislan, Wedge, and
  Debrota]{rogers2006computer}
M.~K. Rogers, J.~Goldman, R.~Mislan, T.~Wedge, and S.~Debrota, ``Computer
  forensics field triage process model,'' in \emph{Proceedings of the
  conference on Digital Forensics, Security and Law}.\hskip 1em plus 0.5em
  minus 0.4em\relax Association of Digital Forensics, Security and Law, 2006,
  p.~27.

\end{thebibliography}

\end{document}